# A Possibilistic and Probabilistic Approach to Precautionary Saving

Irina Georgescu[1]
Adolfo Cristóbal-Campoamor[2]
Ana Mª Lucia-Casademunt[3]

**Summary:** This paper proposes two mixed models to study a consumer's optimal saving in the presence of two types of risk: income risk and background risk. In the first model, income risk is represented by a fuzzy number and background risk by a random variable. In the second model, income risk is represented by a random variable and background risk by a fuzzy number. For each model, three notions of precautionary savings are defined as indicators of the extra saving induced by income and background risk on the consumer's optimal choice. In conclusion, we can characterize the conditions that allow for extra saving relative to optimal saving under certainty, even when a certain component of risk is modelled using fuzzy numbers.

Currently, concern about the high saving rates of Chinese households is widely expressed, as this phenomenon has far-reaching international implications for China's current account surpluses and Western external deficits (Alan Greenspan 2009). Many analysts and policy makers are trying to trace the reasons for this behavior and to predict how long it will last. In this regard, many explanations of this phenomenon point to economic uncertainty as a determinant of precautionary saving (Olivier Blanchard and Francesco Giavazzi 2005; Marcos Chamón, Kai Liu, and Eswar Prasad 2013). In contrast to the Chinese example, European and U.S. households are well known for their low saving rates, which calls for some precise explanation as well (João Sousa Andrade and Adelaide Duarte 2011).

Therefore, it is very important to understand the characteristics of such uncertainty in order to model the household's optimal consumption and saving behavior. Both the kind of uncertainty and the features of consumers' preferences must be analyzed in order to predict micro- or macroeconomic saving patterns. In this context, the notion of precautionary saving has long appeared in models of economic decision under uncertainty. It measures how adding a source of risk modifies optimal saving.

---

[1] Bucharest Academy of Economic Studies, Department of Computer Science and Cybernetics, Bucharest, Romania: irina.georgescu@csie.ase.ro
[2] Universidad Loyola Andalucía, Department of Economics, Sevilla, Spain: acristobal@uloyola.es
[3] Corresponding author. Universidad Loyola Andalucía, Department of Business Organization, Campus de Córdoba, Spain: alucia@uloyola.es

In real life, humans succeed by using imprecise rather than precise knowledge. According to classical logic, however, an extremely deep understanding of the environment is needed to make rational decisions, using exact equations and precise numeric values. In this paper, we attempt to apply fuzzy logic to decision making regarding consumption and saving, incorporating an alternative way of mapping subjective concepts, such as "about right" or "a long time ago", onto exact numeric ranges, according to Robert Fullér (1998). We believe that the connection between fuzzy logic and economic theory needs to be addressed, since it may render different predictions for real life phenomena and different recommended policy measures. Specifically, we will derive the precise conditions for the emergence of precautionary saving when the risk faced by consumers can be modelled not only with random variables but also possibilistically with fuzzy numbers. In this sense, we will be able to quantify in an alternative way the magnitude of the precautionary saving that arises in response to small risks. This information could be useful for authorities seeking to determine the optimal exposure of households to different types of risk in order to encourage or discourage the population to save based oncurrent macroeconomic needs.

Several authors (e.g., Neil A. Doherty and Harris Schlesinger 1983; Christian Gollier and John W. Pratt 1996; Pratt 1998) have studied economic decision processes governed by two types of risk: primary risk (income risk) and background risk (e.g., loss of employment, divorce, illness). In our paper, as in Mario Menegatti (2009), the presence of background risk will be associated with a nonfinancial variable and will be uninsurable, while nevertheless influencing the optimal solution for economic decisions (Louis Eeckhoudt, Gollier, and Schlesinger 2005).

The interaction of both types of risk with optimal saving has been studied by Christophe Courbage and Béatrice Rey (2007) and Menegatti (2009). These models assume that the consumer's activity takes place during two periods and that both types of risk act during the second period. The presence or absence of one of the two types of risk leads to several possible uncertainty situations, with different corresponding definitions of precautionary saving (Menegatti 2009).

All optimal saving models in the literature are based on probability theory. That is, both primary and background risk are modelled as random variables. However, there are risk situations for which probabilistic models are not appropriate (e.g., for small databases). Lotfi A. Zadeh's (1978) possibility theory offers another way to model some risk situations. Here, risk is modelled with possibility distributions (specifically, with fuzzy numbers), and the well-known probabilistic indicators (e.g., expected value, variance, covariance) are replaced with the corresponding possibilistic indicators.

Due to the complexity of economic and financial phenomena, one can have mixed situations in which some risk parameters should be probabilistically modelled with random variables, whereas other risk parameters should be possibilistically modelled with fuzzy numbers. Thus, we can consider the following four situations: (1) a random variable captures primary risk, and a random variable captures background risk; (2) a fuzzy number captures primary risk, and a fuzzy number captures background risk; (3) a fuzzy number captures primary risk, and a random variable captures background risk; (4) a random variable captures primary risk, and a fuzzy number captures background risk.

The above mentioned probabilistic models consider situation (1). The purpose of this paper is to study precautionary saving motives in situations (3) and (4). In situation (3), the risk situation is described by a mixed vector (*A, X*), whereas in situation (4), the risk situation is described by a mixed vector (*Y, B*), where *A, B* are fuzzy numbers, and *X, Y* are random variables. Let us denote the mixed models described by situation (3) as type I models; we will also denote the mixed models described by situation (4) as type II models. For each of those two types of models (type I and type II), we will define three notions of precautionary saving and investigate the necessary and sufficient conditions for extra saving to arise, after adding primary risk, background risk, or both. The main results of this paper establish those necessary and sufficient conditions, expressed in terms of the third-order partial derivatives of the bidimensional utility function and in terms of the probabilistic and possibilistic variances associated with the mixed vector.

We can offer illustrative examples of type I and type II models in real life. For instance, suppose that a man owns a house near a beach surrounded by a wheat field that needs to be harvested every year. If there is less rainfall during the year, the income obtained from the wheat harvest will be lower. The value of renting a room in his house, however, could be higher. Now, assume that only the risk related to the harvest is insurable. However, the background risk is related to potential lodgers' pleasure from sunbathing and could be described by a fuzzy number: "the weather near the beach is pleasant enough for sunbathing". These risks would be negatively correlated, since dry and sunny weather implies that the owner would collect little revenue from the harvest but could probably charge higher rents to his lodgers. Therefore, there may or may not be precautionary saving in response to those two risks. That would be an example of a mixed model of type II. Imagine instead that the main income source for the owner is apartment rent. That rent could be now insurable, unlike the harvest revenue. In this case, income risk would be characterized by a fuzzy number with a value dependent on lodgers' imprecise perceptions of the weather. The situation could be then understood as a mixed model of type I. The objective of this paper is to explore the conditions for precautionary saving in situations such as these two described cases.

We will describe now briefly the structure of the paper. Section 1 provides a historical perspective on the contributions addressing the unidimensional precautionary saving problem, along with probabilistic models of precautionary saving with background risk. Section 2 recalls the indicators of fuzzy numbers and the mixed expected utility of Irina Georgescu and Jani Kinnunen (2011). Section 3 presents the mathematical framework in which the optimal saving models with background risk are embedded. Section 4 proposes mixed models of optimal saving of type I, with income risk modelled as a fuzzy number and background risk modelled as a random variable. Section 5 addresses mixed models of optimal saving of type II, with an income risk modelled with a random variable and a background risk modelled with a fuzzy number. Three notions of precautionary saving will be defined for both mixed models of type I and II. The main results of the paper establish the necessary and sufficient conditions for positive precautionary saving for each notion in terms of the consumer's preferences and the features of risks.

**1. Literature Review**

Saving under uncertainty is introduced in economic research in Hayne E. Leland (1968) and Agmar Sandmo (1970). These papers investigate how the presence of risk modifies the optimal amount of saving. Variation in optimal saving based on risk elements is measured in Leland (1968) and Sandmo (1970) by the concept of precautionary saving. A central result of those papers is that a positive third-order derivative of a consumer's utility function is a necessary and sufficient condition for a risk to increase saving. Miles S. Kimball (1990) defines "prudence" as the sensitivity of optimal saving to the magnitude of risk and introduces an index of absolute prudence as a measure of this sensitivity. He proves that the absolute prudence index is isomorphic to the Arrow-Pratt index of risk aversion Eeckhoudt, Gollier, and Schlesinger (2005). Furthermore, Eeckhoudt and Schlesinger (2008) show that higher-order risk changes affect the demand for saving.

The above papers examine the optimal saving decision in the presence of a unidimensional risk. In particular, the paper by Christopher D. Carroll and Kimball (2008) surveys the literature on univariate precautionary saving. On the other hand, there are also economic and financial situations that require several risk parameters. Therefore, some economic models consider, apart from primary (income) risk, a second kind of risk (exogenous and unhedgeable) usually called "background risk" (see, e.g., Doherty and Schlesinger 1983; Gollier and Pratt 1996; Pratt 1988).

The issue of how the existence of several risk parameters affects the optimal amount of saving thus naturally arose. Courbage and Rey (2007), Eeckhoudt, Rey, and Schlesinger (2007), Menegatti (2009), Diego Nocetti and William T. Smith (2011) investigate the effect of primary and background risk on the optimal saving decision, considering a consumer with a bidimensional utility function. Specifically, the papers by Courbage and Rey (2007) and Menegatti (2009) established the necessary and sufficient conditions for positive precautionary saving to appear in a bivariate context. In particular, Menegatti (2009) defined two notions of precautionary saving: the first explores the impact of a small income risk on saving relative to optimal saving under certainty; the second explores the impact of both sources of risk (primary and background) on saving, also relative to optimal saving under certainty. Moreover, the paper by Nocetti and Smith (2011) develops a precautionary saving model with an infinite horizon. Finally, the paper by Elyés Jouini, Clotilde Napp, and Nocetti (2013) develops a matrix-measure concept of prudence.

Our theoretical setting could be extended in different ways, some of which could encompass alternative saving motives other than the precautionary motive. Our work could be connected to the literature on status goods and social norms (e.g., Harold L. Cole, George J. Mailath, and Andrew Postlewaite 1992; Richmond Harbaugh 1996; Ed Hopkins and Tatiana Kornienko 2004). When allowing certain goods (e.g., income or savings) to offer utility beyond their consumption value, the pattern of saving behavior could be modified in order to attain a certain status. In this respect, our model can help introduce similar status concerns by means of our background and income risks, as captured by our probabilistic and possibilistic setting.

In particular, the Chinese case has attracted considerable attention in the literature. Shang-Jin Wei and Xiaobo Zhang (2011) suggest that in the context of an imbalance in the share of women within the population, as in contemporary China, people could save "in order to improve their relative standing in the marriage market". They could empirically confirm that Chinese parents with sons increased their saving in order to improve their sons' relative attractiveness for marriage. Moreover, such pressure to save spills over into other households *via* prices in the housing market.

Qingyuan Du and Wei (2013) present a related model in which men and women derive utility from an income variable that is affected by their saving decisions and an "emotional utility" (or "love") variable that is affected by their wife or husband. The latter was modelled as a random variable following a probabilistic process such that every couple was formed according to some "emotional utility" threshold above which certain men would be accepted by certain women. Their framework does not include income uncertainty, and the values of the "emotional utility" variable are perfectly revealed before marriage. In calibrations of their model, they showed that a 15% increase in the sex ratio "would lead to an increase in the aggregate private savings […] of about 30-60% of the actual increase in the household savings rate observed in the [Chinese] data".

Alternatively, love or emotional utility may be more appropriately modelled by fuzzy numbers. That may be because people make these decisions in a diffuse environment in which a partner could be fuzzily considered "*about right (or wrong)*" to marry with based on wealth and love. Therefore, our contribution can help add realism to similar decision-making processes, which then better fit the actual phenomena they try to describe. Moreover, in our setting, income and background (love) risks are modelled together, which combines the precautionary saving motive with the competitive saving motive in the same theoretical framework.

Furthermore, we could extend our model naturally to incorporate the general equilibrium effect of pre-marital behavior on aggregate savings as the exogenous sex ratio is modified in the economy. To do so, we would need to incorporate the matching of men and women in the marriage market. Therefore, our model is amenable to very relevant both theoretical and empirical real world applications and is useful as a first step before delving into the specific modelling of these phenomena.

## 2. Preliminaries

### 2.1 Preliminaries on Fuzzy Numbers

Let us now introduce some preliminary concepts in fuzzy theory. They will be useful in setting up the consumer's optimization problem in which one of the sources of risk is represented by a fuzzy number.

Let $X$ be a non-empty set of states. A fuzzy subset of $X$ is a function $A: X \to [0,1]$. A fuzzy set $A$ is normal if $A(x) = 1$ for some $x \in X$. The support of $A$ is defined by $\sup p(A) = \{x \in X \mid A(x) > 0\}$.

Next, assume that $X = \mathbf{R}$. For $\gamma \in [0,1]$, the $\gamma$-level set $[A]^\gamma$ of $A$ is defined by

$$[A]^\gamma = \begin{cases} \{x \in R \mid A(x) \geq \gamma\} & if \quad \gamma > 0 \\ cl(\sup p(A)) & if \quad \gamma = 0 \end{cases}$$

($cl(\sup p(A))$ is the topological closure of $\sup p(A)$).

The fuzzy set $A$ is fuzzy convex if $[A]^\gamma$ is a convex subset of **R** for all $\gamma \in [0,1]$.

A fuzzy subset $A$ of **R** is called a fuzzy number if it is normal, fuzzy convex, and continuous, with bounded support. If $A, B$ are fuzzy numbers and $\lambda \in \mathbf{R}$, then the fuzzy numbers $A+B$ and $\lambda A$ are defined by

$$(A+B)(x) = \sup_{y+z=x} \min(A(y), B(z))$$

$$(\lambda A)(x) = \sup_{\lambda y = x} A(y)$$

A non-negative and monotone increasing function $f:[0,1] \to R$ is a weighting function iff $\int_0^1 f(\gamma) d\gamma = 1$.

We fix a weighting function $f$ and a fuzzy number $A$ such that $[A]^\gamma = [a_1(\gamma), a_2(\gamma)]$ for $\gamma \in [0,1]$. Let $u: R \to R$ be a continuous function (interpreted as a utility function).

The possibilistic expected utility $E(f, u(A))$ is defined by:

$$E(f, u(A)) = \frac{1}{2} \int_0^1 [u(a_1(\gamma)) + u(a_2(\gamma))] f(\gamma) d\gamma. \tag{1}$$

If $u$ is the identity function, then from (1), one obtains the $f$-weighted possibilistic expected value $E(f, A)$ (Fullér and Peter Majlender 2003):

$$E(f, A) = \frac{1}{2} \int_0^1 [a_1(\gamma) + a_2(\gamma)] f(\gamma) d\gamma \tag{2}$$

If $u(x) = (x - E(f, A))^2$, then one obtains the $f$-weighted possibilistic variance (Wei-Guo Zhang and Ying-Luo Wang 2007):

$$Var(f, A) = \frac{1}{2} \int_0^1 [(a_1(\gamma) - E(f, A))^2 + (a_2(\gamma) - E(f, A))^2] f(\gamma) d\gamma. \tag{3}$$

When $f(\gamma) = 2\gamma$ for $\gamma \in [0,1]$, $E(f, A)$ and $Var(f, A)$ are the possibilistic mean value and the possibilistic variance of Christer Carlsson and Fullér (2001), respectively.

## 2.2 Mixed Expected Utilities

The concept of mixed expected utility was introduced by Georgescu and Kinnunen (2011) in order to build a model of risk aversion with mixed parameters: some of them were described by fuzzy numbers and others by random variables. This same notion has been used by Georgescu (2012b) to study mixed investment models in the presence of background risk.

In this section, we will review this definition of mixed expected utility and some of its properties. For clarity, we will consider only the bidimensional case. Then, a mixed vector will have the form $(A, X)$ where $A$ is a fuzzy number, and $X$ is a random variable. Without loss of generality, we only consider the case $(A, X)$.

Let $X$ be a random variable w.r.t. a probability space $(\Omega, K, P)$. We will denote by $M(X)$ its expected value and by $Var(X)$ its variance. If $u: R \to R$ is a continuous function, then $u(X) = u \circ X$ is a random variable, and $M(u(X))$ is the (probabilistic) expected utility of $X$ w.r.t. $u$.

We choose a fixed weighting function $f$ and a bidimensional, continuous utility function $u: R^2 \to R$. Let $(A, X)$ be our particular mixed vector. Assume that the level sets of $A$ have the form $[A]^\gamma = [a_1(\gamma), a_2(\gamma)], \gamma \in [0,1]$. For any $u \in R$, $u(a, X): \Omega \to R$ will be the random variable defined by $u(a, X)(w) = u(a, X(w))$ for any $w \in \Omega$.

Let us now define our concept of mixed expected utility, which will be subject to maximization in our approach to optimal saving.

**Definition 2.1** (Georgescu and Kinnunen 2011; Georgescu 2012a)

*The mixed expected utility $E(f, u(A, X))$ associated with $f, u$ and the mixed vector $(A, X)$ is defined by*

$$E(f, u(A, X)) = \frac{1}{2} \int_0^1 [M(u(a_1(\gamma), X)) + M(u(a_2(\gamma), X))] f(\gamma) d\gamma \quad (4)$$

**Remark 2.2**

(i) *If the fuzzy number $A$ is the constant $a$, then $E(f, u(A, X)) = M(u(a, X))$.*

(ii) If the random variable $X$ is the constant $b$ then

$E(f, u(A, b)) = \frac{1}{2} \int_0^1 [u(a_1(\gamma), b) + u(a_2(\gamma), b)] f(\gamma) d\gamma$.

The following two propositions are essential for proving the main theorems discussed in the subsequent sections:

**Proposition 2.3** (Georgescu and Kinnunen 2011; Georgescu 2012a)

*Let $g, h$ be two bidimensional utility functions and $a, b \in R$. If $u = ag + bh$, then $E(f, u(A, X)) = aE(f, g(A, X)) + bE(f, h(A, X))$.*

**Proposition 2.4** (Georgescu and Kinnunen 2011; Georgescu 2012a)

*If the utility function $u$ has the form $u(x, y) = (x - E(f, A))(y - M(X))$, then $E(f, u(A, X)) = 0$.*

## 3. A Probabilistic Approach to Optimal Saving

The optimal saving models presented in Courbage and Rey (2007) and Menegatti (2009) consider the existence of two types of risk, background risk and income risk, both of which are mathematically represented by random variables. In this section, we will present the general features of these models as a benchmark to start building our main models in the following sections.

The two-period models proposed by Courbage and Rey (2007) and Menegatti (2009) are characterized as follows:

- $u(y, x)$ and $v(y, x)$ are consumer utility functions for period 0 and 1, respectively;

- $y$ represents income, and $x$ is a non-financial variable;
- for period 0 the variables $y$ and $x$ have the certain values $y_0$ and $x_0$, respectively;
- for period 1, there is an uncertain income (described by the random variable $Y$) and a background risk (described by the random variable $X$).

Here, $\bar{y} = M(Y)$ and $\bar{x} = M(X)$. In Menegatti (2009), there are four possible situations for the variables $y$ and $x$:

(a) $y = Y$, $x = X$ (simultaneous presence of income risk and background risk);
(b) $y = Y$, $x = \bar{x}$ (income risk and no background risk);
(c) $y = \bar{y}$, $x = X$ (background risk and no income risk);
(d) $y = \bar{y}$, $x = \bar{x}$ (no uncertainty).

Consider the following expected lifetime utilities corresponding to the situations (a), (c) and (d), respectively:

$$V(s) = u(y_0 - s, x_0) + M(v(Y + s, X)) \tag{5}$$

$$W(s) = u(y_0 - s, x_0) + M(v(\bar{y} + s, X)) \tag{6}$$

$$T(s) = u(y_0 - s, x_0) + v(\bar{y} + s, \bar{x}) \tag{7}$$

where $s$ is the level of saving. According to Menegatti (2009), the optimization problem can be formulated as follows:

$$\max_s V(s) = V(s^*) \tag{8}$$

$$\max_s W(s) = W(s^\circ) \tag{9}$$

$$\max_s T(s) = T(s^{\circ\circ}) \tag{10}$$

with the optimal solutions $s^* = s^*(Y, X)$, $s^\circ = s^\circ(\bar{y}, X)$ and $s^{\circ\circ} = s^{\circ\circ}(\bar{y}, \bar{x})$.

The differences $s^* - s^\circ$ and $s^* - s^{\circ\circ}$ are called *precautionary saving* and *two-source precautionary saving*, respectively, in Menegatti (2009). Finally, the author presents some necessary and sufficient conditions such that $s^* - s^\circ \geq 0$ and $s^* - s^{\circ\circ} \geq 0$, which generalize the results previously obtained by Courbage and Rey (2007).

In this paper, we intend to offer an alternative setting to Menegatti's (2009) by allowing for the possibility that either income risk or background risk is formally represented by a fuzzy number. Therefore, the structure of the following sections parallels Menegatti (2009), although instead of his random vector $(Y, X)$, we will consider mixed vectors of the type $(A, X)$ or $(Y, B)$.

## 4. Mixed Models of Type I

The mixed models of this section are based on the hypothesis that income risk is described by a fuzzy number $A$ and background risk is described by a random variable $X$. We will preserve the notation introduced in the previous section. A fuzzy number $A$ corresponds to the variable $y$, and the random variable $X$ corresponds to the variable $x$. Thus instead of Menegatti's (2009) random vector $(Y, X)$, we have a mixed vector $(A, X)$.

We will fix a weighting function $f$, $a = E(f, A)$ and $\bar{x} = M(X)$. In this case, situations (a)-(d) in Section 3 become:

($a_1$) $y = A$, $x = X$

($b_1$) $y = A$, $x = \bar{x}$

($c_1$) $y = a$, $x = X$

($d_1$) $y = a$, $x = \bar{x}$.

In this section, we will examine how optimal saving changes as we follow the routes $(c_1) \to (a_1)$, $(d_1) \to (a_1)$ and $(b_1) \to (a_1)$. The first two are analogous to the cases studied in Menegatti (2009) for probabilistic models. We will define three notions of precautionary saving and establish necessary and sufficient conditions for these indicators to be positive.

Assume that the bidimensional utility functions $u$ and $v$ are strictly increasing with respect to each component, strictly concave and three times continuously differentiable. Here, $u_i$, $u_{ij}$, $u_{ijk}$ ($v_i$, $v_{ij}$, $v_{ijk}$) denote the first, second and third partial derivatives of $u$ ($v$), respectively.

Next, we will use, as in Menegatti (2009), the following Taylor approximation:

$$v_1(y+s, x) \approx v_1(a+s, \bar{x}) + v_{11}(a+s, \bar{x})(y-a) + v_{12}(a+s, \bar{x})(x-\bar{x}) +$$
$$+ \frac{1}{2}[v_{111}(a+s, \bar{x})(y-a)^2 + v_{122}(a+s, \bar{x})(x-\bar{x})^2 + 2v_{112}(a+s, \bar{x})(y-a)(x-\bar{x})]$$

(11)

Using the notion of mixed expected utility from Subsection 2.2, we introduce the following expected lifetime utilities:

$$V_1(s) = u(y_0 - s, x_0) + E(f, v(A+s, X)) \tag{12}$$

$$W_1(s) = u(y_0 - s, x_0) + E(f, v(a+s, X)) = u(y_0 - s, x_0) + M(v(a+s, X)) \tag{13}$$

$$T_1(s) = u(y_0 - s, x_0) + v(a+s, \bar{x}) \tag{14}$$

$$U_1(s) = u(y_0 - s, x_0) + E(f, v(A+s, \bar{x})). \tag{15}$$

Here, $V_1, W_1, T_1$ are the analogues of $V, W, T$ respectively, in Section 3, and $U_1$ comes from situation ($b_1$) above. Taking into account Formula (1) from Subsection 2.1:

$$V_1(s) = u(y_0 - s, x) + \tag{16}$$
$$+ \frac{1}{2}\int_0^1 [M(v(a_1(\gamma)+s, X)) + M(v(a_2(\gamma)+s, X))]f(\gamma)d\gamma.$$

If we differentiate, from (16) one obtains:

$$V_1'(s) = -u_1(y_0 - s, x_0) +$$

$$+\frac{1}{2}\int_0^1 [M(v_1(a_1(\gamma)+s,X))+M(v_1(a_2(\gamma)+s,X))]f(\gamma)d\gamma$$

which, by Formula (1) in Subsection 2.1, can be written as:

$$V_1'(s)=-u_1(y_0-s,x_0)+E(f,v_1(A+s,X)). \qquad (17)$$

If we differentiate, from (13)-(15) it follows that:

$$W_1'(s)=-u_1(y_0-s,x_0)+M(v_1(a+s,X)) \qquad (18)$$

$$T_1'(s)=-u_1(y_0-s,x_0)+v_1(a+s,\bar{x}) \qquad (19)$$

$$U_1'(s)=-u_1(y_0-s,x_0)+E(f,v_1(A+s,\bar{x})). \qquad (20)$$

**Proposition 4.1** *The functions $V_1, W_1, T_1$ and $U_1$ are strictly concave.*

**Proof.** Differentiating the expression for $V_1'(s)$ above:

$$V_1''(s)=u_{11}(y_0-s,x_0)+$$

$$+\frac{1}{2}\int_0^1 [M(v_{11}(a_1(\gamma)+A,X))+M(v_{11}(a_2(\gamma)+A,X))]f(\gamma)d\gamma.$$

Since $u_{11}<0$ and $v_{11}<0$, one obtains $V_1''(s)<0$. Similarly, one can prove that the other three functions are strictly concave. This ends the proof.

We consider now the following optimization problems:

$$\max_s V_1(s)=V_1(s_1^*) \qquad (21)$$

$$\max_s W_1(s)=W_1(s_1^\circ) \qquad (22)$$

$$\max_s T_1(s)=T_1(s_1^{\circ\circ}) \qquad (23)$$

$$\max_s U_1(s)=U_1(s_1^\Delta) \qquad (24)$$

in which $s_1^*=s_1^*(A,X)$, $s_1^\circ=s_1^\circ(a,X)$, $s_1^{\circ\circ}=s_1^{\circ\circ}(a,\bar{x})$ and $s_1^\Delta=s_1^\Delta(A,\bar{x})$ are optimal solutions.

By Proposition 4.1, the four optimal solutions are given by:

$$V_1'(s_1^*)=0, \ W_1'(s_1^\circ)=0, \ T_1'(s^{\circ\circ})=0, \ U_1'(s_1^\Delta)=0.$$

Taking into account (17)-(20), the optimal conditions are written:

$$u_1(y_0-s_1^*,x_0)=E(f,v_1(A+s_1^*,X)) \qquad (25)$$

$$u_1(y_0-s_1^\circ,x_0)=M(v_1(a+s_1^\circ,X)) \qquad (26)$$

$$u_1(y_0-s_1^{\circ\circ},x_0)=M(v_1(a+s_1^{\circ\circ},X))=v_1(a+s_1^{\circ\circ},\bar{x}) \qquad (27)$$

$$u_1(y_0-s_1^\Delta,x_0)=E(f,v_1(A+s_1^\Delta,\bar{x})). \qquad (28)$$

Following Menegatti (2009), we introduce three notions of mixed precautionary saving: $s_1^*-s_1^\circ$, $s_1^*-s_1^{\circ\circ}$, $s_1^*-s_1^\Delta$.

Here, $s_1^*-s_1^\circ$ corresponds to precautionary saving in Menegatti (2009) and measures the change in optimal saving when moving from $(y=a, x=X)$ to $(y=A, x=X)$, i.e., by adding income risk $A$ in the presence of background risk $X$. The difference $s_1^* - s_1^{\infty}$ expresses the change in optimal saving when moving from the certainty situation $(y = a, x = \bar{x})$ to $(y = A, x = X)$, i.e., by adding income risk $A$ and background risk $X$. Finally, $s_1^* - s_1^{\Delta}$ measures the change in optimal saving when moving from $(y = A, x = \bar{x})$ to $(y = A, x = X)$, i.e., by adding background risk $X$ in the presence of income risk $A$.

Next, we intend to provide necessary and sufficient conditions for the three indicators to be positive.

**Proposition 4.2** *Let $(A, X)$ be a mixed vector with $a = E(f, A)$ and $\bar{x} = M(X)$. The following inequalities are equivalent:*

(i) $s_1^*(A, X) - s_1^{\circ}(a, X) \geq 0$;

(ii) $v_{111}(a + s_1^*(A, x), \bar{x}) \geq 0$.

The intuition behind this result is connected to the usual emergence of precautionary saving in probabilistic models. The optimality of the consumer's decision requires the first period marginal utility to be equal to the second period *expected* marginal utility of the consumption good. If the marginal utility function is convex, then the expected marginal utility in the second period will rise in the presence of a mean-preserving spread. Consequently, the consumption in the first period will need to fall. Therefore, precautionary saving will be the optimal response to risk if and only if the third derivative of the utility function is positive, i.e., if the marginal utility curve is convex.

**Proof.** Using the approximation Formula (11) and Proposition 2.3, by applying the mixed expected utility operator one obtains:

$$E(f, v_1(A + s, X)) \approx v_1(a + s, \bar{X}) +$$
$$+ v_{11}(a + s, \bar{x})E(f, A - a) + v_{12}(a + s, \bar{x})M(X - \bar{x}) +$$
$$+ \frac{1}{2}[v_{111}(a + s, \bar{x})E(f, (A - a)^2) + v_{122}(a + s, \bar{x})M((x - \bar{x})^2) +$$
$$+ 2v_{112}(a + s, \bar{x})E(f, (A - a)(X - \bar{x}))].$$

Note that $E(f, A - a) = 0$, $M(X - \bar{x}) = 0$, $E(f, (A - a)^2) = Var(f, A)$ and $M((X - \bar{x})^2) = Var(X)$. Additionally, $E(f, (A - a)(X - \bar{x})) = 0$; thus, the previous relation becomes:

$$E(f, v_1(A + s, X)) \approx v_1(a + s, \bar{x}) + \frac{1}{2}v_{111}(a + s, \bar{x})Var(f, A) +$$

(29)

$$+ \frac{1}{2}v_{122}(a + s, \bar{x})Var(X).$$

A similar computation shows that:

$$E(f, v_1(a + s, X)) \approx v_1(a + s, \bar{x}) + \frac{1}{2}v_{122}(a + s, \bar{x})Var(X). \tag{30}$$

Taking into account that $W_1$ is strictly concave, it follows that $W_1'$ is strictly decreasing; therefore, $s_1^* \geq s_1^{\circ}$ iff $W_1'(s_1^*) \leq W_1'(s_1^{\circ})$. By (18) and (25):

$$W_1'(s_1^*) = -u_1(y_0 - s_1^*, x_0) + M(v_1(a + s_1^*, X))$$
$$= M(v_1(a + s_1^*, X)) - E(f, v_1(A + s_1^*, X)).$$

By approximating $M(v_1(a + s_1^*, X))$ and $E(f, v_1(A + s_1^*, X))$ with the values given by formulas (29) and (30), from the previous relation one obtains:

$$W_1'(s_1^*) \approx -v_{111}(a + s_1^*, \bar{x})Var(f, A). \tag{31}$$

However, $Var(f, A) \geq 0$; thus, $s_1^* \geq s_1^\circ$ if $W_1'(s_1^*) \leq 0$ iff $v_{111}(a + s_1^*, \bar{x}) \geq 0$.

This ends the proof.

Property (i) of the previous proposition says that the effect of adding income risk $A$ in the presence of background risk $X$ is an increase in optimal saving. In particular, from Proposition 4.2, it follows that if $v_{111} > 0$, then $s_1^*(A, X) - s_1^\circ(a, X) \geq 0$ for any income risk $A$ and any background risk $X$.

Next, we study the change in optimal saving on the route $(d_1) \to (a_1)$.

**Proposition 4.3** *Let $(A, X)$ be a mixed vector with $a = E(f, A)$ and $\bar{x} = M(X)$. The following are equivalent:*

(i) $s_1^*(A, X) - s_1^{\infty}(a, \bar{x}) \geq 0$

(ii) $v_{111}(a + s_1^*(A, X), \bar{x})Var(f, A) + v_{122}(a + s_1^*(A, X), \bar{x})Var(X) \geq 0$.

In this case, both income and background risk have an impact on the second-period expected marginal utility of consumption. Therefore, the convexity of the marginal utility of consumption must be evaluated with respect to the whole mixed vector.

**Proof.** Taking into account that $T_1$ is strictly concave, $s_1^* \geq s_1^{\infty}$ iff $T_1'(s_1^*) \leq T_1'(s_1^{\infty}) = 0$.

By the Formulas (19) and (23) one has the equalities:
$$T_1'(s_1^*) = -u_1(y_0 - s_1^*, x_0) + v_1(a + s_1^*, \bar{x})$$
$$= v_1(a + s_1^*, \bar{x}) - E(f, v_1(A + s_1^*, X)).$$

Formula (21) gives the following approximation:
$$E(f, v_1(A + s_1^*, X)) \approx v_1(a + s_1^*, \bar{x}) + \frac{1}{2}v_{111}(a + s_1^*, \bar{x})Var(f, A) +$$
$$+ \frac{1}{2}v_{122}(a + s_1^*, \bar{x})Var(X),$$

thus,
$$T_1'(s_1^*) \approx -\frac{1}{2}[v_{111}(a + s_1^*, \bar{x})Var(f, A) + v_{122}(a + s_1^*, \bar{x})Var(X)].$$

The following equivalences follow:

$s_1^\circ \geq s_1^{\infty}$ iff $T_1'(s_1^*) \leq 0$

iff $v_{111}(a + s_1^*, \bar{x})Var(f, A) + v_{122}(a + s_1^*, \bar{x})Var(X) \geq 0$.

This ends the proof.

Condition (i) of Proposition 4.3 says that the effect of adding income risk $A$ and background risk $X$ is an increase in optimal saving. In particular, from Proposition 4.3, it follows that if $v_{111} \geq 0$ and $v_{122} \geq 0$, then for any mixed vector $(A, X)$, $s_1^* - s_1^{\infty} \geq 0$.

**Corollary 4.4** *Assume that $(A, X)$ is a mixed vector and $s_1^*(A, X) - s_1^\circ(a, X) \geq 0$. If $v_{122} \geq 0$ then $s_1^*(A, X) - s_1^{\infty}(a, \bar{x}) \geq 0$, where $a = E(f, A)$ and $\bar{x} = M(X)$.*

**Proof.** The proof follows from Propositions 4.2 and 4.3, taking into account that $Var(f, A) \geq 0$ and $Var(X) \geq 0$. This ends the proof.

Finally, consider the change in optimal saving on the route $(b_1) \to (a_1)$.

**Proposition 4.5** *Let $(A, X)$ be a mixed vector with $a = E(f, A)$ and $\bar{x} = M(X)$. The following are equivalent:*
(i) $s_1^*(A, X) - s_1^\Delta(A, \bar{x}) \geq 0$;
(ii) $v_{122}(a + s_1^*(A, X), \bar{x}) \geq 0$.

**Proof.** Using approximation formula (11) and applying Proposition 2.3, it follows:
$$E(f, v_1(A + s, \bar{x})) \approx v_1(a + s, \bar{x}) + \frac{1}{2} v_{111}(a + s, \bar{x}) Var(f, A). \tag{32}$$
By (20) and (25):
$$U_1'(s_1^*) = -u_1(y_0 - s_1^*, x_0) + E(f, v_1(A + s_1^*, \bar{x}))$$
$$= E(f, v_1(A + s_1^*, \bar{x})) - E(f, v_1(A + s_1^*, X)).$$
By replacing $E(f, v_1(A + s_1^*, \bar{x}))$ and $E(f, v_1(A + s_1^*, X))$ with their approximate values obtained by applying (32) and (29) one obtains:
$$U_1'(s_1^*) \approx -\frac{1}{2} v_{122}(a + s_1^*, \bar{x}) Var(X).$$
Since $U_1$ is strictly concave, $s_1^* \geq s_1^\Delta$ iff $U_1'(s_1^*) \leq U_1'(s_1^\Delta) = 0$ iff $v_{122}(a + s_1^*, \bar{x}) \geq 0$.
This ends the proof.

Condition (i) of the previous proposition says that adding background risk $X$ in the presence of income risk $A$ leads to an increase in optimal saving.

**Corollary 4.6** *If $s_1^* - s_1^\circ \geq 0$ and $s_1^* - s_1^\Delta \geq 0$, then $s_1^* - s_1^{\infty} \geq 0$.*

**Proof.** Propositions 4.2, 4.3 and 4.5 are applied.

In the next example, we will show that there exist mixed vectors $(A, X)$ with $a = E(f, A)$ and $\bar{x} = M(X)$ and utility functions $u, v$ for which condition $s_1^*(A, X) - s_1^\circ(a, X) \geq 0$ does not imply:

$$s_1^*(A, X) - s_1^\infty(a, \bar{x}) \geq 0.$$

**Example 4.7** Let $c, d$ be two real numbers such that $0 < c < d$. Let $A$ be a fuzzy number with $a_1(\gamma) = c, a_2(\gamma) = d$, for any $\gamma \in [0,1]$ and $X$ the uniform repartition on $[c, d]$.
It is known that $M(X) = \dfrac{c+d}{2}$ and $Var(X) = \dfrac{(c-d)^2}{12}$. A simple calculation shows that $E(f, A) = \dfrac{c+d}{2}$ and $Var(f, A) = \dfrac{(c-d)^2}{4}$. Then:

$$v_{111}(y, x)Var(f, A) + v_{122}(y, x)Var(X) = \frac{(c-d)^2}{4}[v_{111}(y, x) + \frac{1}{3}v_{122}(y, x)]. \tag{33}$$

Assume that the utility function $v$ has the form:

$$v(y, x) = -\frac{1}{\alpha}e^{-\alpha y}\frac{x^{1-\gamma}}{1-\gamma} \quad \text{with } y \in \mathbf{R},\ x > 0, \gamma > 0, \gamma \neq 1$$

One notices that:

$$v_{111}(y, x) = \alpha^2 e^{-\alpha y}\frac{x^{1-\gamma}}{1-\gamma};\ v_{122}(y, x) = -\gamma e^{-\alpha y} x^{-\gamma-1}.$$

Then, from (33) it follows:

$$v_{111}(y, x)Var(f, A) + v_{122}(y, x)Var(X) = \frac{(c-d)^2}{4}e^{-\alpha y}[\frac{x^{1-\gamma}}{1-\gamma}\alpha^2 - \frac{\gamma}{3}x^{-\gamma-1}] \tag{34}$$

One notices:

$$\frac{x^{1-\gamma}}{1-\gamma} - \frac{\gamma}{3}x^{-\gamma-1} \geq 0 \text{ iff } \frac{x^{1-\gamma}}{1-\gamma}\alpha^2 \geq \frac{\gamma}{3}x^{-\gamma-1}$$

$$\text{iff } \frac{x^2}{1-\gamma}\alpha^2 \geq \frac{\gamma}{3}$$

$$\text{iff } \frac{1}{\gamma(1-\gamma)}\alpha^2 \geq \frac{1}{3x^2}.$$

From (34) and these equivalences, it follows:

$$v_{111}(y, x)Var(f, A) + v_{122}(y, x)Var(X) \geq 0 \text{ iff } \frac{\alpha^2}{\gamma(1-\gamma)} \geq \frac{1}{3(c+d)^2} \tag{35}$$

Replacing in (35) $y = a + s_1^*(A, X)$ and $x = \bar{x} = \dfrac{c+d}{2}$ and taking into account Proposition 4.3, one obtains:

$$s_1^*(A, X) - s_1^\infty(a, \bar{x}) \geq 0 \text{ iff } \frac{\alpha^2}{\gamma(1-\gamma)} \geq \frac{4}{3(c+d)^2}. \tag{36}$$

For $\gamma = \dfrac{3}{4}$ from (36), it follows:

$$s_1^*(A, X) - s_1^\infty(a, \bar{x}) \geq 0 \text{ iff } 4\alpha^2 \geq \frac{1}{(c+d)^2} \text{ iff } c + d \geq \frac{1}{2\alpha}.$$

Then, if $c + d < \dfrac{1}{2\alpha}$, we will have $s_1^*(A, X) - s_1^{\infty}(a, \bar{x}) < 0$. On the other hand:

$$y_{111}(y, x) = \alpha^2 e^{-\alpha y} \dfrac{x^{1-\gamma}}{1-\gamma} = \alpha^2 e^{-\alpha y} \dfrac{x^{\frac{1}{4}}}{\frac{1}{4}} = 4\alpha^2 e^{-\alpha y} x^{\frac{1}{4}} < 0$$

Thus, by Proposition 4.2, $s_1^*(A, X) - s_1^{\circ}(a, X) \geq 0$.

The above example shows that the converse of Corollary 4.6 is not true. In particular, we observe that in our example, the income variable $y$ a "good" and the background variable $x$ is a "bad" for the consumer. Specifically, since $\gamma < 1$, the marginal utility of consumption increases at a decreasing rate as $x$ rises, whereas it decreases at a decreasing rate as $y$ rises. As a result, the expected marginal utility of consumption may not be convex in the mixed vector as a whole. This implies that, under both sources of risk, the consumer may not save more than in the certainty situation.

## 5. Mixed Models of Type II

The mixed models in this section assume that income risk is represented by a random variable $Y$ and that background risk is represented by a fuzzy number $B$. We keep the hypotheses in Section 4 on the bidimensional utility functions $u$ and $v$.

We fix a weighting function $f$. Here, $M(y) = \bar{y}$ and $E(f, B) = b$. As in previous sections, we consider the following cases:

($a_2$) $y = Y, x = B$

($b_2$) $y = Y, x = b$

($c_2$) $y = \bar{y}, x = B$

($d_2$) $y = \bar{y}, x = b$.

We analyse how optimal saving changes on the following three routes: $(c_2) \to (a_2)$, $(d_2) \to (a_2)$ and $(b_2) \to (a_2)$. For each of these three cases, we will introduce a notion of precautionary saving, and we will prove the necessary and sufficient conditions for these three indicators to be positive.

Corresponding to cases $(a_2) - (d_2)$, we introduce four expected lifetime utilities:

$$V_2(s) = u(y_0 - s, x_0) + E(f, v(Y + s, B))$$

(37)

$$W_2(s) = u(y_0 - s, x_0) + E(f, v(\bar{y} + s, B)) \tag{38}$$

$$T_2(s) = u(y_0 - s, x_0) + v(\bar{y} + s, b) \tag{39}$$

$$U_2(s) = u(y_0 - s, x_0) + E(f, v(Y + s, b)). \tag{40}$$

Deriving (37)-(40) it follows:

$$V_2'(s) = -u_1(y_0 - s, x_0) + E(f, v_1(Y + s, B)) \tag{41}$$

$$W_2'(s) = -u_1(y_0 - s, x_0) + E(f, v_1(\bar{y} + s, B)) \tag{42}$$

$$T_2'(s) = -u_1(y_0 - s, x_0) + v_1(\bar{y} + s, b) \tag{43}$$

$$U_2'(s) = -u_1(y_0 - s, x_0) + E(f, v_1(Y + s, b)). \tag{44}$$

As in the previous section, it is proved that $V_2, W_2, T_2$ and $U_2$ are strictly concave functions. We form four maximization problems:

$$\max_s V_2(s) = V_2(s_2^*) \tag{45}$$

$$\max_s W_2(s) = W_2(s_2^\circ) \tag{46}$$

$$\max_s T_2(s) = T_2(s_2^{\circ\circ}) \tag{47}$$

$$\max_s U_2(s) = U_2(s_2^\Delta)$$

(48)

in which $s_2^* = s_2^*(Y, B)$, $s_2^\circ = s_2^\circ(\bar{y}, B)$, $s_2^{\circ\circ} = s_2^{\circ\circ}(\bar{y}, b)$ and $s_2^\Delta = s_2^\Delta(Y, b)$ are the optimal solutions.

By (41)-(44), the optimal conditions $V_2'(s_2^*) = 0$, $W_2'(s_2^\circ) = 0$, $T_2'(s_2^{\circ\circ}) = 0$ and $U_2'(s_2^\Delta) = 0$ will be written:

$$u_1(y_0 - s_2^*, x_0) = E(f, v_1(Y + s_2^*, B)) \tag{49}$$

$$u_1(y_0 - s_2^\circ, x_0) = E(f, v_1(\bar{y} + s_2^\circ, B)) \tag{50}$$

$$u_1(y_0 - s_2^{\circ\circ}, x_0) = v_1(\bar{y} + s_1^{\circ\circ}, b) \tag{51}$$

$$u_1(y_0 - s_2^\Delta, x_0) = E(f, v_1(Y + s_2^\Delta, b)). \tag{52}$$

We consider the following notions of precautionary saving: $s_2^* - s_2^0$, $s_2^* - s_2^{\circ\circ}$ and $s_2^* - s_2^\Delta$. The precautionary saving $s_2^* - s_2^0$ measures the change in optimal saving on route $(c_2) \to (a_2)$; $s_2^* - s_2^{\circ\circ}$, changes on route $(d_2) \to (a_2)$; and $s_2^* - s_2^\Delta$, on route $(b_2) \to (a_2)$.

The following three propositions offer the necessary and sufficient conditions for these three indicators to be positive.

**Proposition 5.1** *Let $(Y, B)$ be a mixed vector with $\bar{y} = M(Y)$ and $b = E(f, B)$. The following are equivalent:*

(i) $\quad s_2^*(Y, B) - s_2^\circ(\bar{y}, B) \geq 0$

(ii) $\quad v_{111}(\bar{y} + s_2^*(Y, B), b) \geq 0$.

Sketch of the proof. By applying approximation formula (11) of Section 4 and Propositions 2.3 and 2.4 as in the proof of Proposition 4.5, one reaches:

$$E(f, v_1(Y + s, B)) \approx v_1(\bar{y} + s, b) + \frac{1}{2} v_{111}(\bar{y} + s, b) Var(Y) + \tag{53}$$

$$+ \frac{1}{2} v_{122}(\bar{y} + s, b) Var(f, B)$$

$$E(f, v_1(\bar{y} + s, B)) \approx v_1(\bar{y} + s, b) + \frac{1}{2} v_{122}(\bar{y} + s, b) Var(f, B). \tag{54}$$

By (42) and (49), one obtains:
$$W_2'(s_2^*) = E(f, v_1(\bar{y} + s_2^*, B)) - E(f, v_1(Y + s_2^*, B)). \qquad (55)$$

By replacing in (55) $E(f, v_1(\bar{y} + s_2^*, B))$ and $E(f, v_1(Y + s_2^*, B))$ with their approximate values from (54) and (55), we find the solution:
$$W_2(s_2^*) \approx -\frac{1}{2} v_{111}(\bar{y} + s_2^*, b) Var(Y). \qquad (56)$$

By an analogous argument as in the proof of Proposition 4.2, $s_2^* \leq s_2^\circ$ iff $v_{111}(\bar{y} + s_2^*, b) \geq 0$.

**Proposition 5.2** Let $(Y, B)$ be a mixed vector with $\bar{y} = M(Y)$ and $b = E(f, B)$. The following are equivalent:
(i) $s_2^*(Y, B) - s_2^{\circ\circ}(\bar{y}, b) \geq 0$;
(ii) $v_{111}(\bar{y} + s_2^*(Y, B), b) Var(Y) + v_{122}(\bar{y} + s_2^*(Y, B), b) Var(f, B) \geq 0$.

**Proposition 5.3** Let $(Y, B)$ be a mixed vector with $\bar{y} = M(Y)$ and $b = E(f, B)$. The following are equivalent:
(i) $s_2^*(Y, B) - s_2^\Delta(Y, b) \geq 0$
(ii) $v_{122}(\bar{y} + s_2^*(Y, B), b) \geq 0$.

**Corollary 5.4** Let $(Y, B)$ be a mixed vector with $\bar{y} = M(Y)$ and $b = E(f, B)$. If $s_2^*(Y, B) - s_2^\circ(\bar{y}, B) \geq 0$ and $s_2^*(Y, B) - s_2^\Delta(Y, b) \geq 0$, then $s_2^*(Y, B) - s_2^{\circ\circ}(\bar{y}, b) \geq 0$.

The proofs of Propositions 5.2 and 5.3 and Corollary 5.4 are similar to those of Propositions 4.3 and 4.5 and Corollary 4.6, respectively.

As in Section 4, one proves that the converse of Corollary 5.4 is not true.

Positive precautionary saving notions $s_2^* - s_2^0$, $s_2^* - s_2^{\circ\circ}$ and $s_2^* - s_2^\Delta$ indicate that optimal saving increases on routes $(c_2) \to (a_2)$, $(d_2) \to (a_2)$ and $(b_2) \to (a_2)$. The above results characterize these conditions in terms of the third-order partial derivatives of $v$.

## 6. Conclusions

Our intention in this paper was to find the conditions for precautionary saving in the presence of both probabilistic and possibilistic sources of risk. We set up a decision-making problem for the consumer and compare saving outcomes in the presence and absence of these small risks. Our findings indicate that the crucial determinants of the magnitude of precautionary saving are the third-order partial derivatives of the utility function and the variances corresponding to the mixed vector. We believe our quantitative findings could be of interest to policy makers, especially in the context of developing countries where consumers are often credit constrained, as they are in our paper, and the main saving motives are more closely connected to risk aversion than to wealth accumulation (see Angus Deaton 1989).

From a macroeconomic perspective, it is true that the Chinese corporate savings rate has also risen sharply in the recent years. However, as noted by Thomas W. Bates, Kathleen M. Kahle, and Rene M. Stulz (2009), corporate savings rates seem to have increased around the world. Therefore, the main differential component of Chinese saving behavior is attributable to households (see Wei and Zhang 2011), which adds potential macroeconomic relevance to our analysis.

The optimal saving models in this paper combine the methods of probability and possibility theory. For mixed two-period models, the optimal saving changes are studied in two cases:

(1) income risk is a fuzzy number and background risk is a random variable;
(2) income risk is a random variable and background risk is a fuzzy number.

For each of these two types of models, three notions of mixed precautionary saving have been introduced. These indicators measure variations in optimal saving as a result of adding income risk in the presence of background risk, adding background risk in the presence of income risk or simultaneously adding income risk and background risk to a certain situation. The main results in the paper describe when the three indicators are positive, which indicates that the mentioned modifications increase the optimal level of saving.

Our results also indicate that when the consumer's environment is fuzzy, there are potentially reasons for extra saving relative to certainty. We have also characterized the conditions for such precautionary saving. This characterization could be useful for predicting the behavior of aggregate saving in response to well-defined risks affecting a country's population. A possible extension of our paper could be a numerical comparison of the differences in saving predicted by our model and by that of Menegatti (2009).

The mixed models in this paper follow parallel the probabilistic model of Menegatti (2009), where both background risk and income risk are random variables. A purely possibilistic optimal saving model in which both income and background risk are fuzzy numbers is left for a future analysis.